\pdfoutput=1

\documentclass[final,5p,twocolumn]{elsarticle}

\usepackage{amssymb}
\usepackage{hyperref}

\biboptions{sort&compress}

\begin{document}

\begin{frontmatter}

\title{Hydrodynamic modeling of $^3$He-Au collisions at $\sqrt{s_{NN}}=200$~GeV}

\author[agh]{Piotr Bo\.zek}
\ead{Piotr.Bozek@fis.agh.edu.pl}

\author[ifj,ujk]{Wojciech Broniowski}
\ead{Wojciech.Broniowski@ifj.edu.pl}

\address[agh]{AGH University of Science and Technology, Faculty of Physics and Applied Computer Science, 
30-059 Krak\'ow, Poland}
\address[ifj]{The H. Niewodnicza\'nski Institute of Nuclear Physics, Polish Academy of Sciences, 31-342 Krak\'ow, Poland}
\address[ujk]{Institute of Physics, Jan Kochanowski University, 25-406 Kielce, Poland}

\date{2 March 2015}

\begin{abstract}
Collective flow and femtoscopy in ultrarelativistic $^3$He-Au collisions are investigated 
within the 3+1-dimensional (3+1D) viscous event-by-event hydrodynamics. We evaluate elliptic and triangular 
flow coefficients as functions of the  transverse momentum. We find the typical long-range ridge structures in the two-particle correlations 
in the relative azimuth and pseudorapidity, in the pseudorapidity directions of both Au and $^3$He. 
We also make predictions for the pionic interferometric radii, which decrease with the transverse momentum of the pion pair.
All features found hint on collectivity of the dynamics of the system formed in $^3$He-Au collisions, with 
hydrodynamics leading to quantitative agreement with the up-to-now released data.
\end{abstract}


\end{frontmatter}

\section{Introduction}

The recent experimental~\cite{CMS:2012qk,Abelev:2012ola,Aad:2012gla,Adare:2013piz} and 
theoretical~\cite{Bozek:2011if,Dusling:2012wy,Bozek:2012gr,Qin:2013bha,Shuryak:2013ke,Bzdak:2013zma,Bozek:2013ska,Werner:2013ipa,Nagle:2013lja,Kozlov:2014fqa,Dumitru:2014yza,Bzdak:2014dia}. 
interest in ultrarelativistic heavy-light nuclear collisions originates from expectations that studies of such system 
may shed light on mechanisms governing the formation of long-range correlations, and thus reveal information on dynamics in the  
earliest phases of the reaction. A successful scenario for heavy-light collisions explored in this paper involves, exactly as in the well understood 
case of two heavy ion collisions, collective dynamics (hydrodynamics, transport models). Collectivity provides in a natural way the shape-flow transmutation: the deformation 
of the initial configuration (ellipticity, triangularity, etc.) is transformed event-by-event into harmonic flow at freeze-out. Moreover, the 
approximate translational symmetry on the initial transverse 
shape along the spatial rapidity direction leads to collimated flow, producing the famous ridge structures, 
i.e., azimuthal correlations between particles with a large pseudorapidity separation. 
Yet another vivid feature of collectivity is the mass ordering of various observables~\cite{Bozek:2013ska,Werner:2013ipa}
 seen in proton-nucleus collisions~\cite{Abelev:2013bla}.

The azimuthal deformation of the fireball in small systems is due to random fluctuations, as in p-Pb collisions, 
or to a combination of fluctuations and the intrinsic deformation of the small projectile, as in d-Au collisions, where 
the large intrinsic separation between the proton and neutron leads to large elliptic flow, predicted in~\cite{Bozek:2011if} and verified experimentally
in~\cite{Adare:2013piz}. Moreover, collisions involving projectiles with an intrinsic 
{\em triangular deformation}, such $^3$He-Au \cite{Nagle:2013lja} or $^{12}$C-Au 
\cite{Broniowski:2013dia}, are particularly interesting, as they probe systems with nontrivial projectile geometry.

In this paper we analyze in detail the predictions of 3+1D viscous hydrodynamics of Ref.~\cite{Bozek:2011ua} for 
$^3$He-Au collisions at $\sqrt{s_{NN}}=200$~GeV, the reaction currently analyzed at RHIC. In a previous paper~\cite{Bozek:2014cya}
we have proposed specific tests of collectivity, based on ratios of flow coefficients evaluated with 4- and 2-particle cumulants. 
In the present work we focus on other, more direct and immediately accessible aspects of the reaction, namely, on the 
elliptic and triangular flow coefficients, $v_2$ and $v_3$, and the femtoscopic radii. We also explore the ridge formation, both on the 
$^3$He-side and Au-side in the pseudorapidity direction. A comparison to preliminary flow data from the PHENIX Collaboration~\cite{phenixnapa} indicates 
a successful description of the reaction within our approach. We also confirm the very recent findings of Romatschke~\cite{Romatschke:2015gxa} for the 
flow coefficients and the interferometric radii. 

Throughout this paper we use the three-phase approach, consisting of 1)~the Glauber~\cite{Czyz:1969jg} 
Monte Carlo simulations of the initial state with GLISSANDO~\cite{Rybczynski:2013yba},
the intermediate 3+1D event-by-event viscous hydrodynamics~\cite{Bozek:2011ua}, and the statistical hadronization at freeze-out simulated with 
THERMINATOR~\cite{Chojnacki:2011hb}. The three-nucleon configurations of the $^3$He nucleus 
are taken from the Green's function Monte Carlo
calculations~\cite{Carlson:1997qn} as provided in~\cite{Loizides:2014vua}. The initial entropy density of the fireball in the transverse plane 
follows from the mixed model~\cite{Kharzeev:2000ph,Back:2001xy}, incorporating the wounded nucleons and an admixture of binary collisions, with 
relative weight $\alpha$. We use $\alpha=0.125$ and $\sigma_{NN}^{\rm inel}=42$~mb with a Gaussian wounding profile. The entropy density 
is constructed as a sum of Gaussians in the transverse plane of width $0.4$~fm,  located at the positions of the wounded nucleons.
The assumed longitudinal profile of the fireball along the space-time rapidity is  different for the left- and right-going 
participant nucleons; we use an approximately linear ansatz for the dependence on the spacetime rapidity (see Ref.~ \cite{Bozek:2010bi}
for details), that fairly well  describes the observed asymmetry of the rapidity spectra in d-Au collisions~\cite{Bialas:2004su}.
  
The centrality class $c=0-5\%$ is selected approximately by using events with the number of wounded nucleons $N_w \ge 25$.
The viscous hydrodynamic evolution is performed with shear viscosity  $\eta/s=0.08$ and bulk viscosity 
$\zeta/s=0.04$ (for $T<170$ MeV). The evolution starts at $\tau=0.6$~fm/c, and ends at the freeze-out 
hypersurface with temperature of 150~MeV.  
The non-equilibrium corrections to the local momentum distributions at freeze-out are implemented in the
THERMINATOR code. Our analysis is carried out with 500 hydrodynamic events, where on top of each we generate 5000 THERMINATOR events.

\section{Flow coefficients and long range ridge correlations \label{sec:flow}}

\begin{figure}[tb]
\begin{center}
\includegraphics[angle=0,width=0.49 \textwidth]{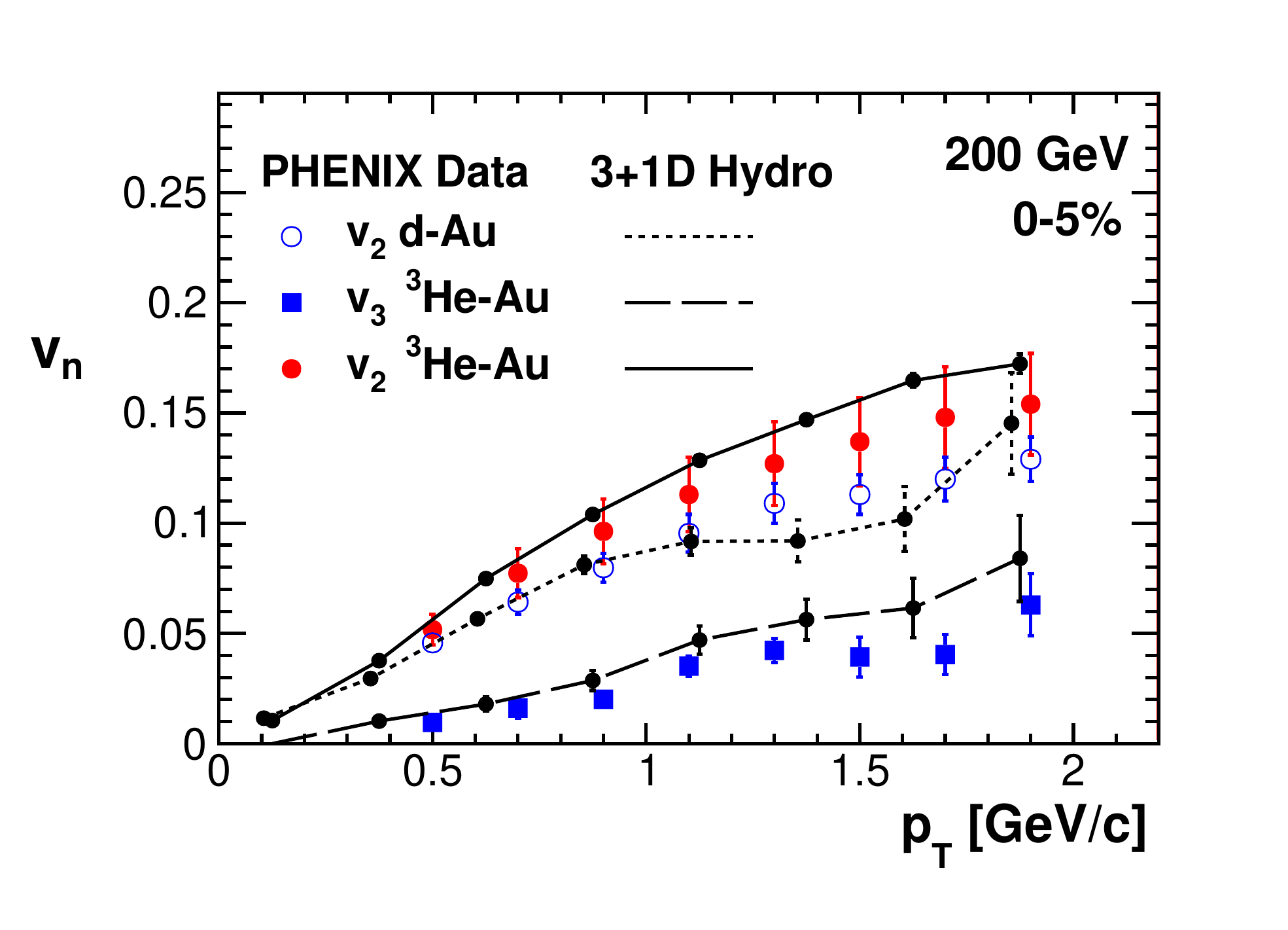}
\end{center}
\vspace{-11mm}
\caption{Elliptic and triangular flow coefficients of charged hadrons as functions of the transverse momentum, $p_T$, for
$^3$He-Au collisions: PHENIX preliminary data~\cite{phenixnapa} (full circles and full squares, respectively)
and hydrodynamic calculations (solid and long-dashed lines respectively), 
as well as elliptic flow coefficient for d-Au collisions: PHENIX data 
\cite{Adare:2014keg} (empty circles) and hydrodynamic 
calculations (dashed line).  
\label{fig:v23heau}} 
\end{figure}  

\begin{figure}[tb]
\begin{center}
\includegraphics[angle=0,width=0.49 \textwidth]{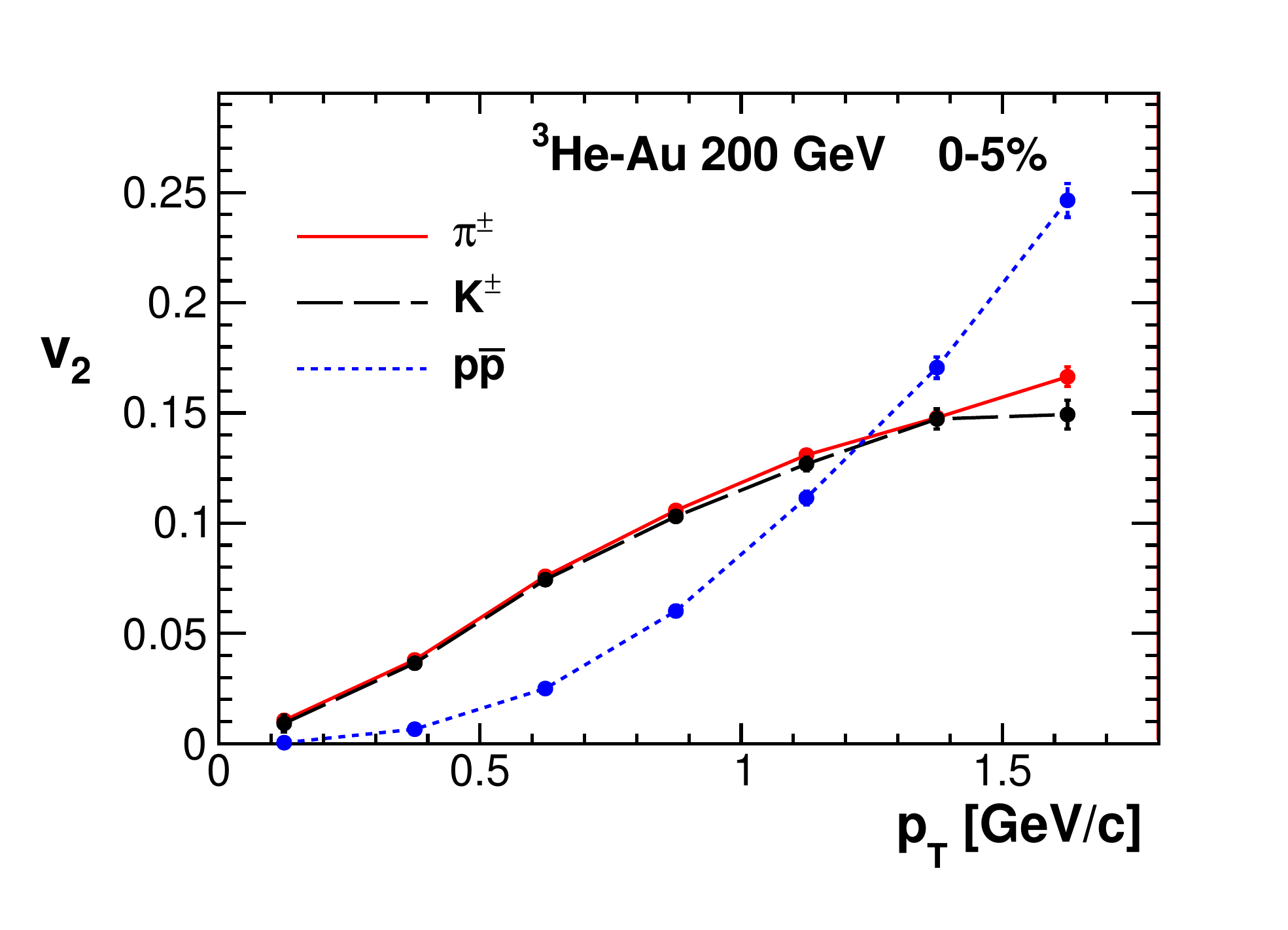}
\end{center}
\vspace{-11mm}
\caption{Elliptic flow coefficient for charged pions, kaons, and protons+antiprotons (solid, long-dashed, and dashed lines respectively) from the 3+1D hydrodynamic calculations, plotted
as functions of the transverse momentum $p_T$.
\label{fig:v2mass}} 
\end{figure} 

We begin the presentation of our results with the elliptic and triangular flow coefficients, evaluated as functions of the transverse momentum, $p_T$,
for all charged hadrons (Fig.~\ref{fig:v23heau}). We evaluate the flow coefficients of charged hadrons with $|\eta|<0.5$,
using the scalar-product method~\cite{Adler:2002pu,Luzum:2012da}, with the reference particles taken from the Au-side bin, $3.1<\eta<3.9$ 
and $0.3< p_\perp < 5.0$~GeV.
We note large values of both coefficients, in approximate agreement with the preliminary results of the PHENIX 
Collaboration~\cite{phenixnapa}. For comparison, we also show $v_2$ for d-Au collisions, which is somewhat smaller 
than in the $^3$He-Au case and compatible with the data. We note that a larger value of $v_2$ in $^3$He-Au 
collisions is expected  from the large initial ellipticity of the fireball~\cite{Bozek:2014cya}.

In Fig.~\ref{fig:v2mass} we show the mass hierarchy of $v_2(p_T)$, plotting it for charged pions, kaons, and protons+antiprotons. The behavior is typical 
for hydrodynamics, with the $p$+$\bar p$ case significantly lower than pions or kaons at low $p_T$, and higher above $p_T\simeq 1.2$~GeV. 

\begin{figure}[tb]
\includegraphics[angle=0,width=0.45 \textwidth]{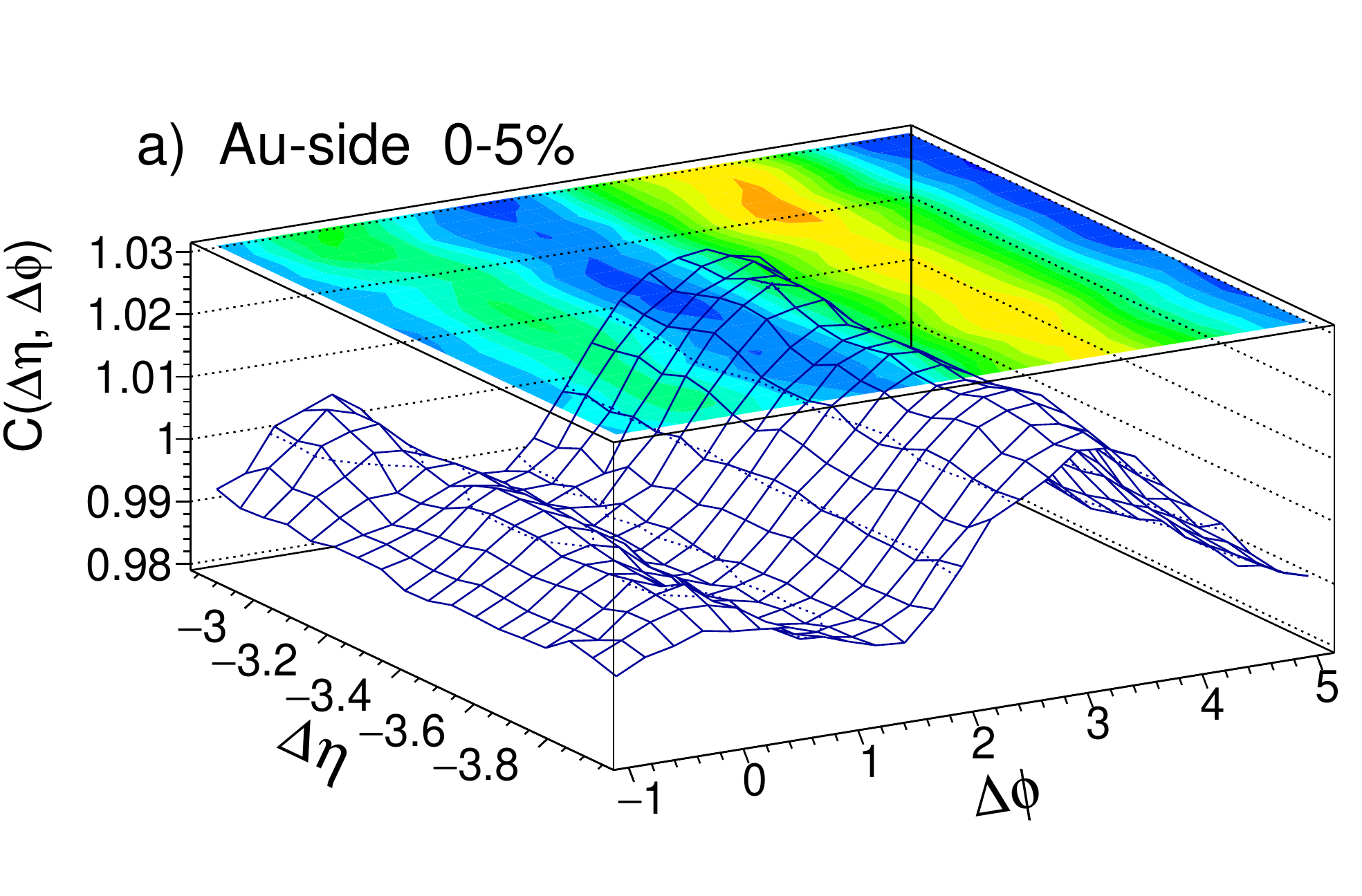} \vspace{-9mm} \\
\includegraphics[angle=0,width=0.45 \textwidth]{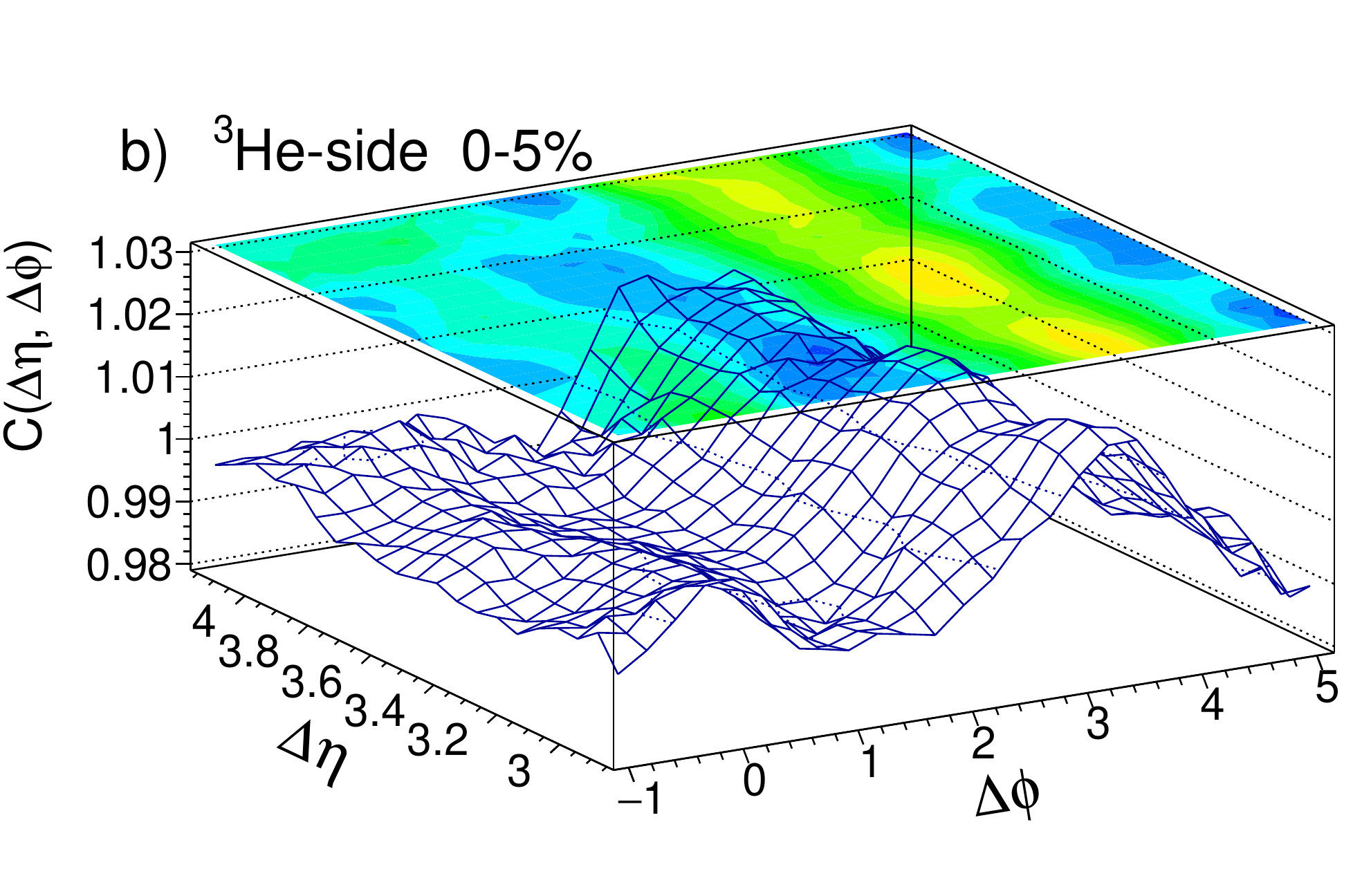}
\vspace{-4mm}
\caption{The two-particle correlation function $C(\Delta\eta,\Delta \phi)$  for a)~Au-side and b)~$^3$He-side. The near- and away-side ridges are clearly visible. 
The surface has been smoothed for better visibility. The kinematic cuts are specified in the text.
\label{fig:ridge0}} 
\end{figure} 

\begin{figure}[tb]
\vspace{-10mm}
\includegraphics[angle=0,width=0.45 \textwidth]{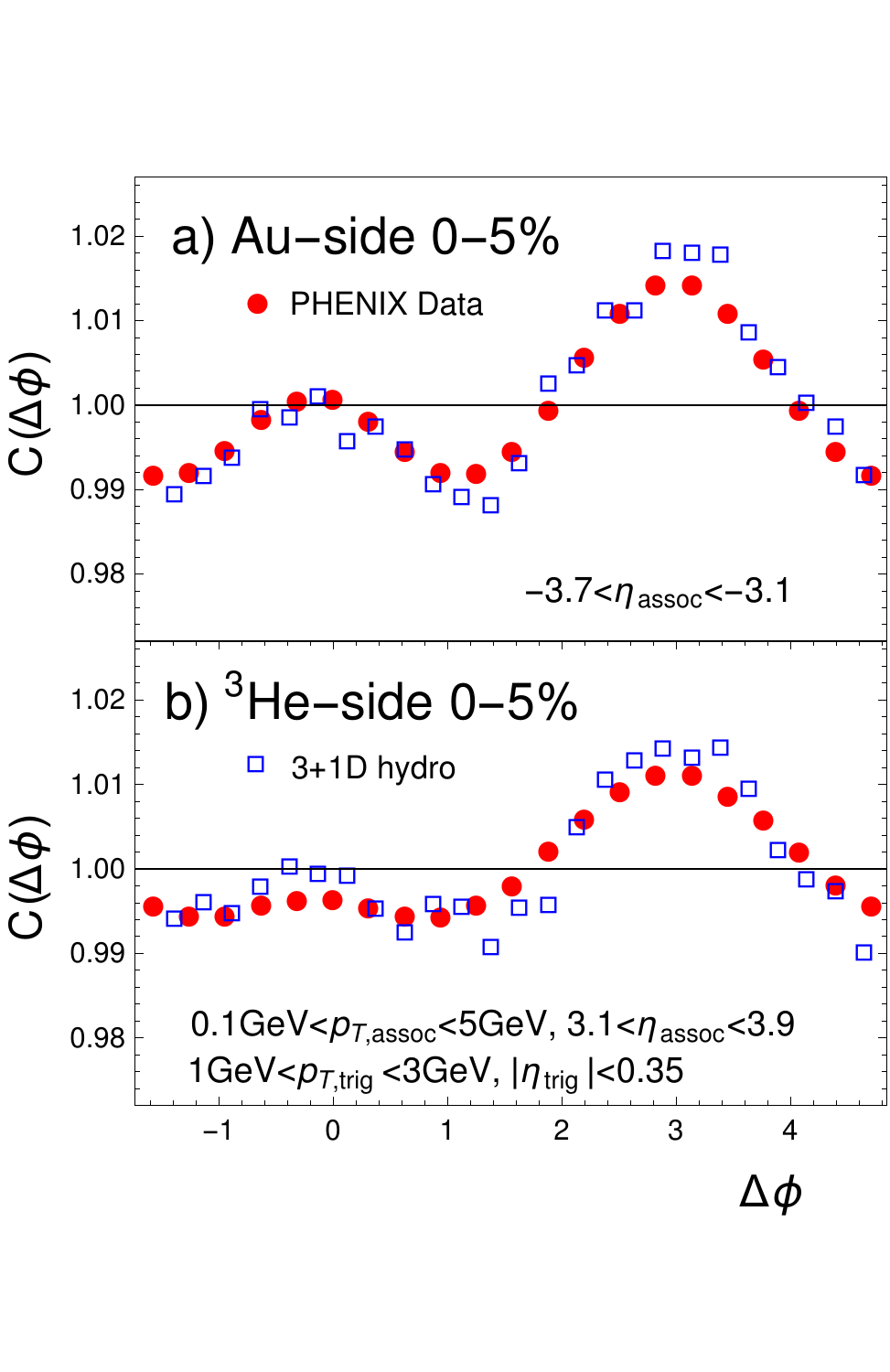}
\vspace{-16mm}
\caption{The two-particle correlation function $C(\Delta \phi)$ for a)~Au-side and b)~$^3$He-side. The ridge structure is clearly visible for both cases in the data and in the model.
The model (empty squares) uses the same kinematic cuts as the PHENIX experiment. The preliminary data (filled dots) come from Ref.~\cite{phenixnapa}.
\label{fig:ridge}} 
\end{figure} 

A very vivid manifestation of collectivity is made with the help of the two-particle correlation functions, defined in the standard way as 
\begin{eqnarray}
C(\Delta\eta,\Delta \phi) &=& \frac{S(\Delta \eta, \Delta \phi)}{B(\Delta \eta, \Delta \phi)}, \nonumber \\
C(\Delta \phi) &=& \frac{\int d{\Delta \eta} \, S(\Delta \eta, \Delta \phi)}{\int d{\Delta \eta} \, B(\Delta \eta, \Delta \phi)}, \label{eq:C}
\end{eqnarray}
where the signal $S$ is constructed by histogramming the pairs of particles with the relative pseudorapidity $\Delta \eta$ and the 
relative azimuth $\Delta \phi$, while the background $B$ is the analogous quantity evaluated with the mixed events. 
The kinematic cuts for the two particles correspond to the PHENIX experiment: \mbox{$|\eta_{\rm trig}| < 0.35$}, \mbox{$1{\rm ~GeV} < p_{T, {\rm  trig}} < 3{\rm ~GeV}$}, and 
\mbox{$-3.7 < \eta_{\rm assoc} < -3.1$} or  \mbox{$3.1 < \eta_{\rm assoc} < 3.9$} for the Au-side and $^3$He-side, respectively. 
The result of our model simulation for $C(\Delta\eta,\Delta\phi)$ are shown in Fig.~\ref{fig:ridge0}. We note the clear appearance of the near- and away-side ridges. 

The result of the projected correlation function $C(\Delta\phi)$, as well as the PHENIX preliminary data~\cite{phenixnapa}, are shown in Fig.~\ref{fig:ridge}. 
Again, we note clearly the emergence of the ridge structure, both in the Au-side and in the $^3$He-side, in fair agreement with the data. 
We remark that since the kinematic cuts are rather narrow, limiting the number of the observed hadrons, it is quite challenging to accumulate enough statistics in the hydrodynamic 
simulation to have statistically significant results.
Qualitatively similar results have been obtained in the 
AMPT model \cite{Koop:2015wea}.

We note that the precise results for $C(\Delta \phi)$ are sensitive to the kinematic cuts mimicking the
detector acceptance and efficiency. Here we have used $p_{T, {\rm assoc}}>0.1$~GeV.  A higher $p_T$ cut would lead to stronger flow, and, consequently, to a higher
ridge amplitude. If better accuracy is desired, Monte Carlo simulations should carefully incorporate the response of the detectors. 

In our THERMINATOR simulations we have incorporated approximately the transverse momentum conservation, which is crucial for the reproduction of the 
features of the correlation functions. We have used a procedure which accepts only those events where the total 
transverse momentum is sufficiently small. We have verified that limiting the acceptance window to $|\sum_i p_{T,i}|<5$~GeV 
is enough, which corresponds to accepting only about 10\% of all events \cite{Bozek:2012gr}. The momentum conservation builds up the strength of the 
away-side ridge, while the near-side ridge comes predominantly as a combination of the second and third harmonic flow components
in the correlation function. It interesting to note that the observed near-side ridge structure for $^3$He-side rapidities can be explained as
an effect of the hydrodynamic expansion of the fireball; the collective
expansion correlates particles at forward and backward rapidities with the event-plane defined  by the geometry of the fireball.

\section{Interferometric radii \label{sec:hbt}}

\begin{figure}
\begin{center}
\vspace{-5mm}
\includegraphics[angle=0,width=0.49 \textwidth]{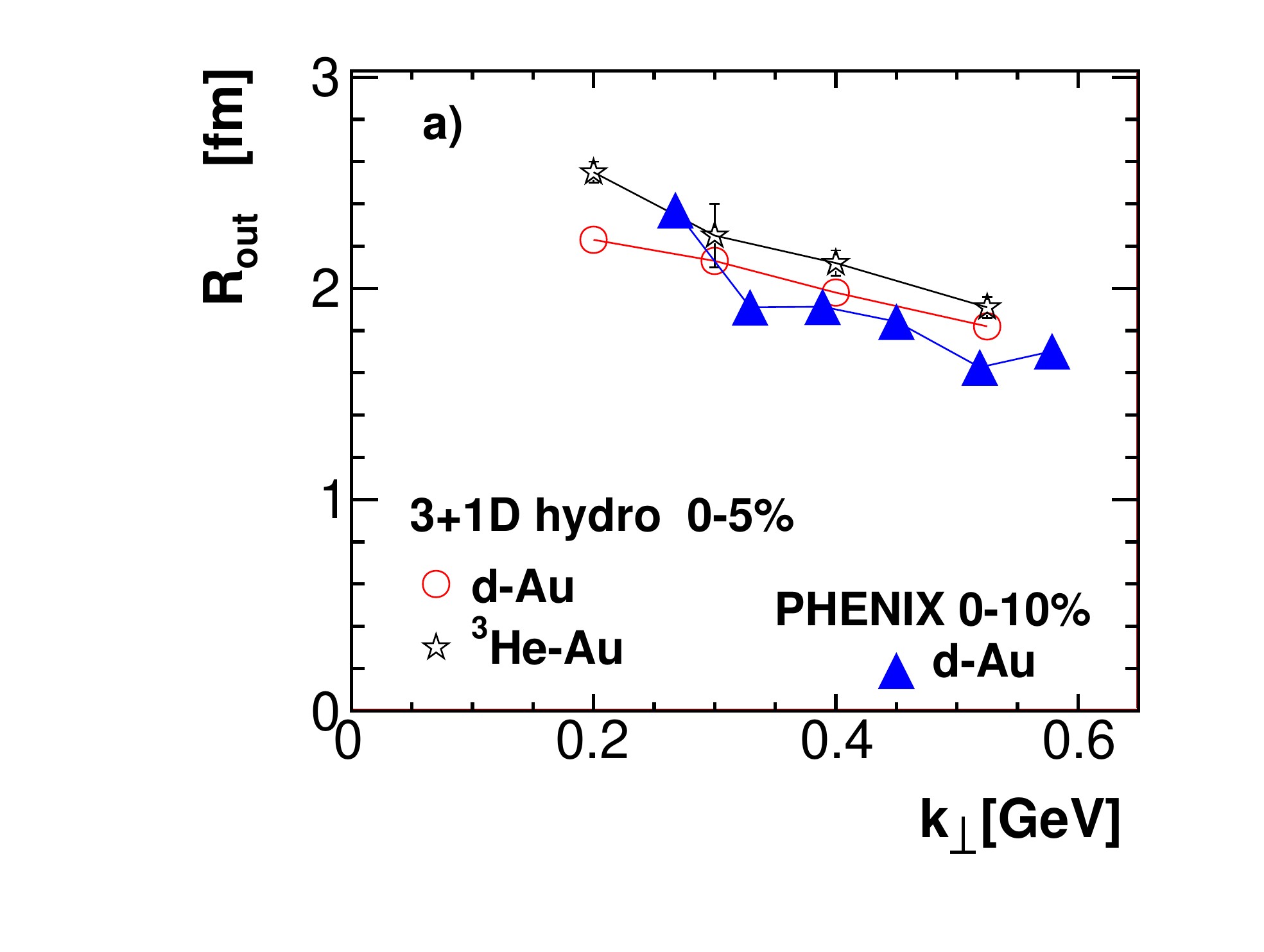}  \vspace{-18mm}\\
\includegraphics[angle=0,width=0.49 \textwidth]{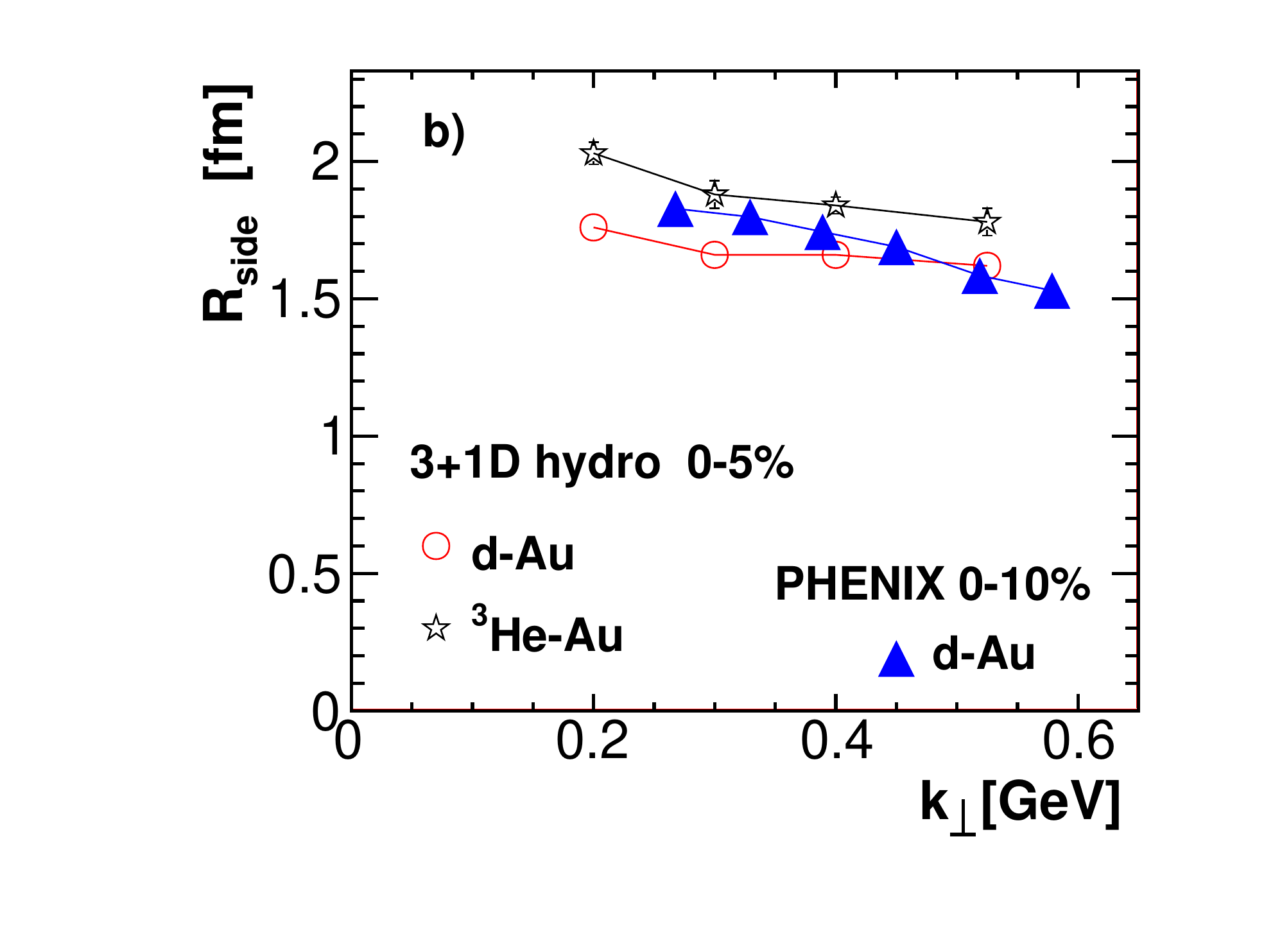}  \vspace{-18mm}\\
\includegraphics[angle=0,width=0.49 \textwidth]{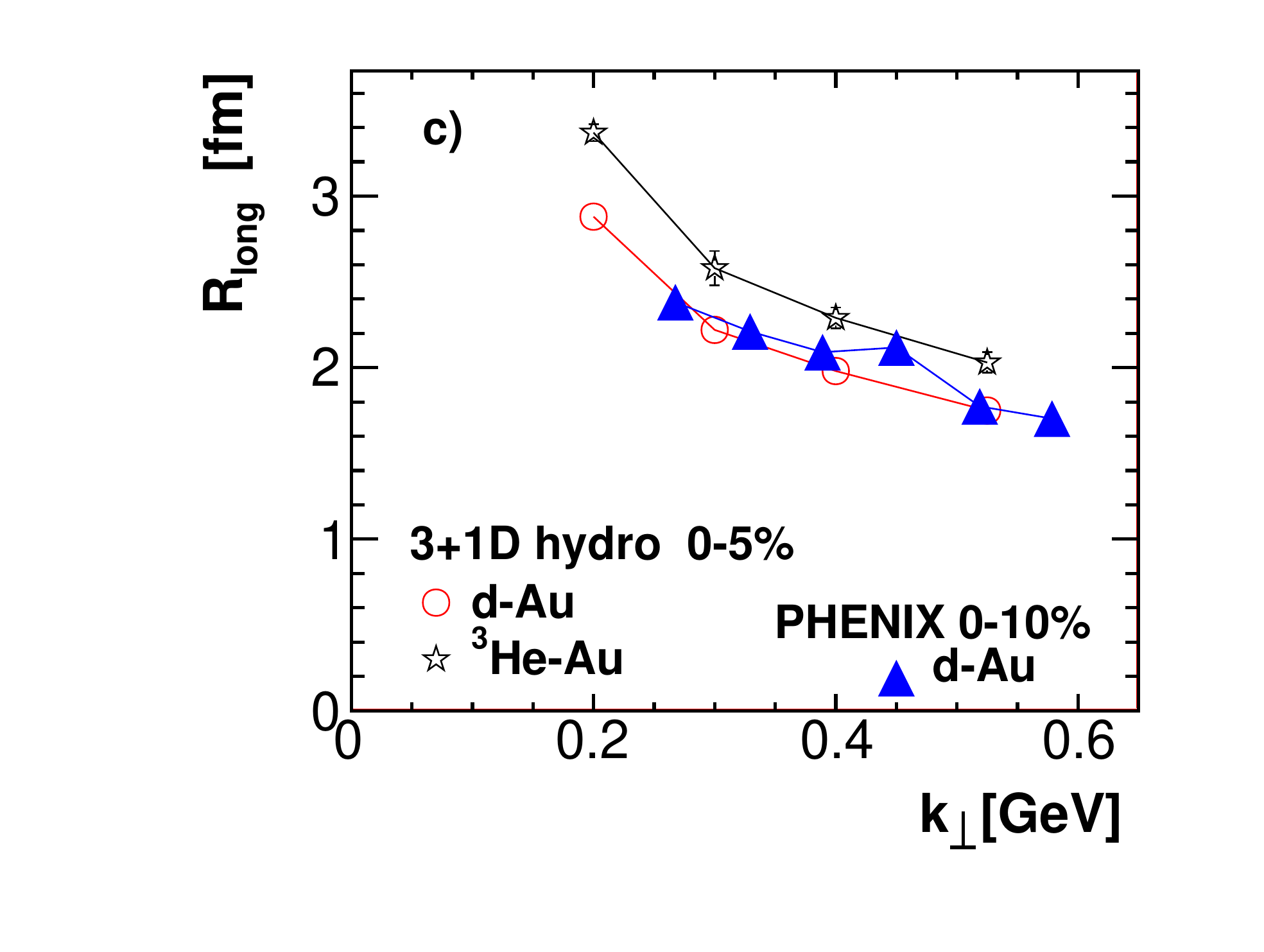}   \vspace{-18mm}\\
\includegraphics[angle=0,width=0.49 \textwidth]{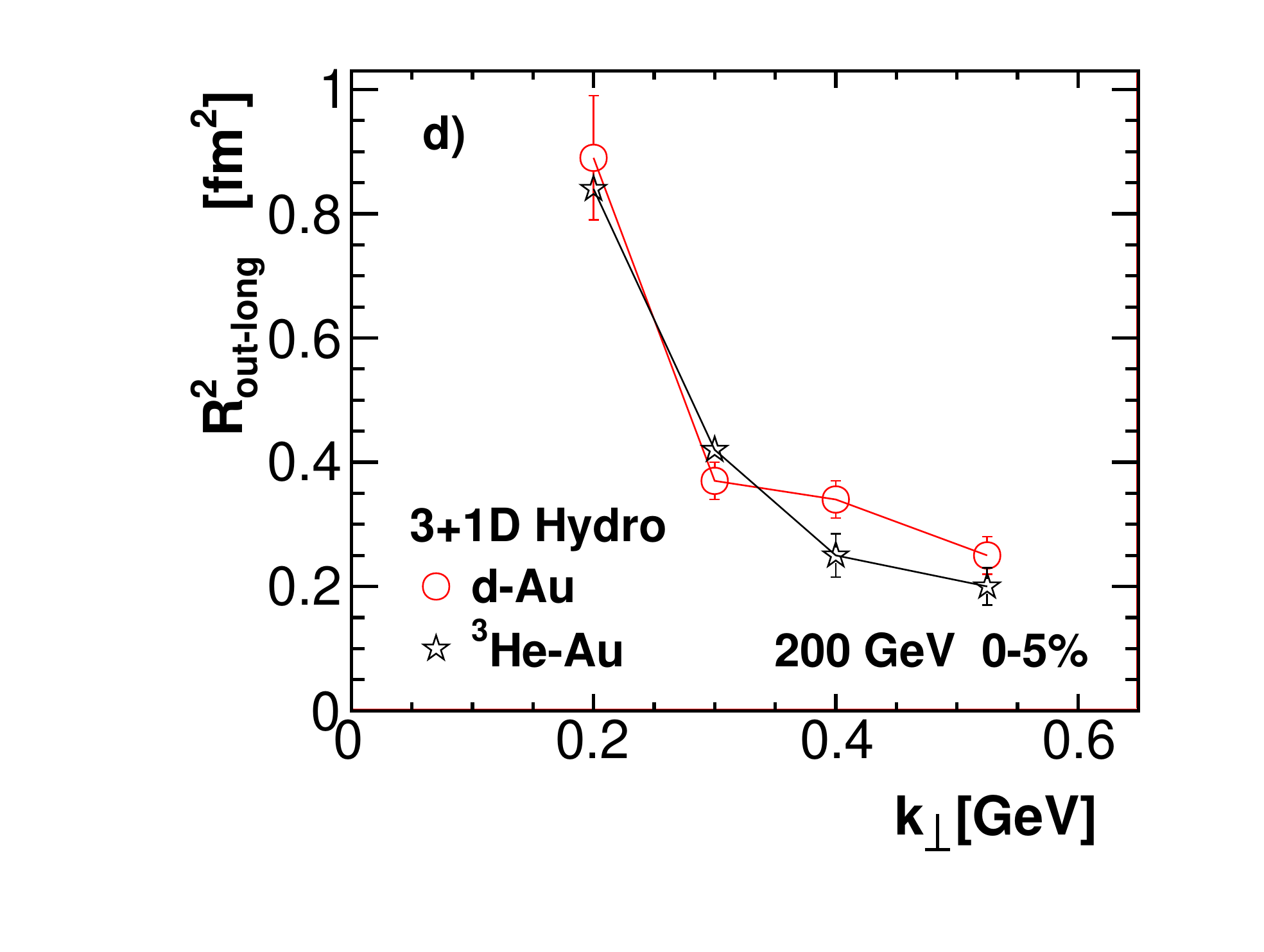} \vspace{-16mm}
\end{center}
\caption{Pionic interferometric radii $R_{\rm out}$, $R_{\rm side}$, $R_{\rm long}$, and
$R^2_{\rm out-long}$ (panels a) through d), respectively), plotted as functions of the average pion pair momentum for $^3$He-Au 
 (stars) and d-Au (circles) collisions. The triangles denote the experimental 
 results of the PHENIX collaboration for the  d-Au system~\cite{Adare:2014vri}.
\label{fig:hbt}} 
\end{figure} 

To obtain the Hanbury Brown-Twiss pionic correlation radii, we use the two-particle method \cite{Kisiel:2006is}
to evaluate the three dimensional 
correlation function in the relative pion momentum ($q_{\rm out}$, $q_{\rm side}$, $q_{\rm long}$). 
For each  given bin of average transverse momentum of the pair, $k_\perp$,
we  construct histograms of  the correlated and mixed pairs from THERMINATOR events~\cite{Chojnacki:2011hb}.
The correlation function, evaluated as a ratio of these histograms, is fitted with Gaussians,
\begin{eqnarray}
&& \hspace{-11mm} C(q,k_\perp)  =  1 +  \\ 
&& \hspace{-7mm} \lambda  e^{-R_{\rm out}^2 q_{\rm out}^2-R_{\rm side}^2q_{\rm side}^2 - 
 R_{\rm long}^2q_{\rm long}^2-2R_{\rm out-long}^2 q_{\rm out}q_{\rm long}}.  \nonumber 
 \label{eq:gauss2} 
\end{eqnarray}
The additional mixed term involving $q_{\rm out}q_{\rm long}$ appears for asymmetric collisions~\cite{Bozek:2014fxa}; including or dropping this
term does not change the fitted values of the radii $R_{\rm out}$, $R_{\rm side}$, and $R_{\rm long}$.
Figure~\ref{fig:hbt} presents our results, with stars indicating the  $^3$He-Au system.
We note that the radii fall down with $k_\perp$, as expected from the presence of flow. 
The fall-off is very fast for the small parameter $R^2_{\rm out-long}$.  

The $k_\perp$ dependence of the interferometric radii in central  d-Au collisions 
(open circles in Fig.~\ref{fig:hbt}) is similar as for \mbox{$^3$He-Au}, and is in reasonable agreement with the PHENIX data~\cite{Adare:2014vri}.
The $^3$He-Au radii are, as expected, somewhat larger than for the d-Au case. 
We note that the ratio of the rms radii of $^3$He and d is 1.14, similar to the corresponding 
ratio for the $R_{\rm side}$ radii.

\section{Conclusion}

We reiterate our main results:

\begin{enumerate}

\item The transverse-momentum dependent elliptic and triangular flow coefficients, $v_2(p_T)$ and $v_3(p_T)$, are 
as in the preliminary data of the PHENIX collaboration~\cite{phenixnapa}.

\item The characteristic ridge structure is found, consistent with the data~\cite{phenixnapa}, both along the Au and the $^3$He sides.

\item We have made predictions for the particle-identified elliptic flow coefficient $v_2(p_T)$, showing the expected mass ordering.

\item We have evaluated the Hanbury-Brown Twiss interferometric radii for pion pairs, 
which are somewhat larger than in the d-Au system. They decrease with the transverse momentum of the pair.

\end{enumerate}

All above features are consistent with the collective interpretation of the fireball dynamics formed in ultrarelativistic nuclear collisions, 
indicating that evolution at all rapidities is driven by the same universal mechanism,
determined by the elliptic and triangular deformation of the initial fireball, supplemented with fluctuations.

\bigskip
                               
Research supported by the Polish Ministry of Science and Higher Education (MNiSW), by the
National Science Center grants DEC-2012/05/B/ST2/02528 and DEC-2012/06/A/ST2/00390, as well as by PL-Grid Infrastructure.

\bigskip

\bibliography{../hydr}

\begin{thebibliography}{38}
\expandafter\ifx\csname natexlab\endcsname\relax\def\natexlab#1{#1}\fi
\providecommand{\url}[1]{\texttt{#1}}
\providecommand{\href}[2]{#2}
\providecommand{\path}[1]{#1}
\providecommand{\DOIprefix}{doi:}
\providecommand{\ArXivprefix}{arXiv:}
\providecommand{\URLprefix}{URL: }
\providecommand{\Pubmedprefix}{pmid:}
\providecommand{\doi}[1]{\href{http://dx.doi.org/#1}{\path{#1}}}
\providecommand{\Pubmed}[1]{\href{pmid:#1}{\path{#1}}}
\providecommand{\bibinfo}[2]{#2}
\ifx\xfnm\relax \def\xfnm[#1]{\unskip,\space#1}\fi
\bibitem[{Chatrchyan et~al.(2013)}]{CMS:2012qk}
\bibinfo{author}{S.~Chatrchyan}, et~al. (\bibinfo{collaboration}{CMS
  Collaboration}), \bibinfo{journal}{Phys. Lett.} \bibinfo{volume}{B718}
  (\bibinfo{year}{2013}) \bibinfo{pages}{795}.
  \href{http://arxiv.org/abs/1210.5482}{arXiv:1210.5482}.
\bibitem[{Abelev et~al.(2013)}]{Abelev:2012ola}
\bibinfo{author}{B.~Abelev}, et~al. (\bibinfo{collaboration}{ALICE
  Collaboration}), \bibinfo{journal}{Phys. Lett.} \bibinfo{volume}{B719}
  (\bibinfo{year}{2013}) \bibinfo{pages}{29}.
  \DOIprefix\doi{10.1016/j.physletb.2013.01.012}.
  \href{http://arxiv.org/abs/1212.2001}{arXiv:1212.2001}.
\bibitem[{Aad et~al.(2013)}]{Aad:2012gla}
\bibinfo{author}{G.~Aad}, et~al. (\bibinfo{collaboration}{ATLAS
  Collaboration}), \bibinfo{journal}{Phys. Rev. Lett.} \bibinfo{volume}{110}
  (\bibinfo{year}{2013}) \bibinfo{pages}{182302}.
  \DOIprefix\doi{10.1103/PhysRevLett.110.182302}.
  \href{http://arxiv.org/abs/1212.5198}{arXiv:1212.5198}.
\bibitem[{Adare et~al.(2013)}]{Adare:2013piz}
\bibinfo{author}{A.~Adare}, et~al. (\bibinfo{collaboration}{PHENIX
  Collaboration}), \bibinfo{journal}{Phys. Rev. Lett.} \bibinfo{volume}{111}
  (\bibinfo{year}{2013}) \bibinfo{pages}{212301}.
  \DOIprefix\doi{10.1103/PhysRevLett.111.212301}.
  \href{http://arxiv.org/abs/1303.1794}{arXiv:1303.1794}.
\bibitem[{Bo\.zek(2012)}]{Bozek:2011if}
\bibinfo{author}{P.~Bo\.zek}, \bibinfo{journal}{Phys. Rev.}
  \bibinfo{volume}{C85} (\bibinfo{year}{2012}) \bibinfo{pages}{014911}.
  \href{http://arxiv.org/abs/1112.0915}{arXiv:1112.0915}.
\bibitem[{Dusling and Venugopalan(2013)}]{Dusling:2012wy}
\bibinfo{author}{K.~Dusling}, \bibinfo{author}{R.~Venugopalan},
  \bibinfo{journal}{Phys. Rev. D} \bibinfo{volume}{87} (\bibinfo{year}{2013})
  \bibinfo{pages}{054014}. \DOIprefix\doi{10.1103/PhysRevD.87.054014}.
  \href{http://arxiv.org/abs/1211.3701}{arXiv:1211.3701}.
\bibitem[{Bo\.zek and Broniowski(2013)}]{Bozek:2012gr}
\bibinfo{author}{P.~Bo\.zek}, \bibinfo{author}{W.~Broniowski},
  \bibinfo{journal}{Phys. Lett.} \bibinfo{volume}{B718} (\bibinfo{year}{2013})
  \bibinfo{pages}{1557}.
  \href{http://arxiv.org/abs/1211.0845}{arXiv:1211.0845}.
\bibitem[{Qin and M{\"u}ller(2014)}]{Qin:2013bha}
\bibinfo{author}{G.-Y. Qin}, \bibinfo{author}{B.~M{\"u}ller},
  \bibinfo{journal}{Phys. Rev.} \bibinfo{volume}{C89} (\bibinfo{year}{2014})
  \bibinfo{pages}{044902}. \DOIprefix\doi{10.1103/PhysRevC.89.044902}.
  \href{http://arxiv.org/abs/1306.3439}{arXiv:1306.3439}.
\bibitem[{Shuryak and Zahed(2013)}]{Shuryak:2013ke}
\bibinfo{author}{E.~Shuryak}, \bibinfo{author}{I.~Zahed},
  \bibinfo{journal}{Phys. Rev.} \bibinfo{volume}{C88} (\bibinfo{year}{2013})
  \bibinfo{pages}{044915}. \DOIprefix\doi{10.1103/PhysRevC.88.044915}.
  \href{http://arxiv.org/abs/1301.4470}{arXiv:1301.4470}.
\bibitem[{Bzdak et~al.(2013)Bzdak, Schenke, Tribedy, and
  Venugopalan}]{Bzdak:2013zma}
\bibinfo{author}{A.~Bzdak}, \bibinfo{author}{B.~Schenke},
  \bibinfo{author}{P.~Tribedy}, \bibinfo{author}{R.~Venugopalan},
  \bibinfo{journal}{Phys. Rev.} \bibinfo{volume}{C87} (\bibinfo{year}{2013})
  \bibinfo{pages}{064906}. \DOIprefix\doi{10.1103/PhysRevC.87.064906}.
  \href{http://arxiv.org/abs/1304.3403}{arXiv:1304.3403}.
\bibitem[{Bo\.zek et~al.(2013)Bo\.zek, Broniowski, and
  Torrieri}]{Bozek:2013ska}
\bibinfo{author}{P.~Bo\.zek}, \bibinfo{author}{W.~Broniowski},
  \bibinfo{author}{G.~Torrieri}, \bibinfo{journal}{Phys. Rev. Lett.}
  \bibinfo{volume}{111} (\bibinfo{year}{2013}) \bibinfo{pages}{172303}.
  \DOIprefix\doi{10.1103/PhysRevLett.111.172303}.
  \href{http://arxiv.org/abs/1307.5060}{arXiv:1307.5060}.
\bibitem[{Werner et~al.(2014)Werner, Bleicher, Guiot, Karpenko, and
  Pierog}]{Werner:2013ipa}
\bibinfo{author}{K.~Werner}, \bibinfo{author}{M.~Bleicher},
  \bibinfo{author}{B.~Guiot}, \bibinfo{author}{I.~Karpenko},
  \bibinfo{author}{T.~Pierog}, \bibinfo{journal}{Phys. Rev. Lett.}
  \bibinfo{volume}{112} (\bibinfo{year}{2014}) \bibinfo{pages}{232301}.
  \DOIprefix\doi{10.1103/PhysRevLett.112.232301}.
  \href{http://arxiv.org/abs/1307.4379}{arXiv:1307.4379}.
\bibitem[{Nagle et~al.(2014)Nagle, Adare, Beckman, Koblesky, Koop
  et~al.}]{Nagle:2013lja}
\bibinfo{author}{J.~Nagle}, \bibinfo{author}{A.~Adare},
  \bibinfo{author}{S.~Beckman}, \bibinfo{author}{T.~Koblesky},
  \bibinfo{author}{J.~O. Koop}, et~al., \bibinfo{journal}{Phys.Rev.Lett.}
  \bibinfo{volume}{113} (\bibinfo{year}{2014}) \bibinfo{pages}{112301}.
  \DOIprefix\doi{10.1103/PhysRevLett.113.112301}.
  \href{http://arxiv.org/abs/1312.4565}{arXiv:1312.4565}.
\bibitem[{Kozlov et~al.(2014)Kozlov, Luzum, Denicol, Jeon, and
  Gale}]{Kozlov:2014fqa}
\bibinfo{author}{I.~Kozlov}, \bibinfo{author}{M.~Luzum},
  \bibinfo{author}{G.~Denicol}, \bibinfo{author}{S.~Jeon},
  \bibinfo{author}{C.~Gale}  (\bibinfo{year}{2014}).
  \href{http://arxiv.org/abs/1405.3976}{arXiv:1405.3976}.
\bibitem[{Dumitru et~al.(2014)Dumitru, McLerran, and Skokov}]{Dumitru:2014yza}
\bibinfo{author}{A.~Dumitru}, \bibinfo{author}{L.~McLerran},
  \bibinfo{author}{V.~Skokov}  (\bibinfo{year}{2014}).
  \href{http://arxiv.org/abs/1410.4844}{arXiv:1410.4844}.
\bibitem[{Bzdak and Ma(2014)}]{Bzdak:2014dia}
\bibinfo{author}{A.~Bzdak}, \bibinfo{author}{G.-L. Ma},
  \bibinfo{journal}{Phys.Rev.Lett.} \bibinfo{volume}{113}
  (\bibinfo{year}{2014}) \bibinfo{pages}{252301}.
  \DOIprefix\doi{10.1103/PhysRevLett.113.252301}.
  \href{http://arxiv.org/abs/1406.2804}{arXiv:1406.2804}.
\bibitem[{Abelev et~al.(2013)}]{Abelev:2013bla}
\bibinfo{author}{B.~B. Abelev}, et~al. (\bibinfo{collaboration}{ALICE
  Collaboration})  (\bibinfo{year}{2013}).
  \href{http://arxiv.org/abs/1307.1094}{arXiv:1307.1094}.
\bibitem[{Broniowski and Arriola(2014)}]{Broniowski:2013dia}
\bibinfo{author}{W.~Broniowski}, \bibinfo{author}{E.~R. Arriola},
  \bibinfo{journal}{Phys. Rev. Lett.} \bibinfo{volume}{112}
  (\bibinfo{year}{2014}) \bibinfo{pages}{112501}.
  \DOIprefix\doi{10.1103/PhysRevLett.112.112501}.
  \href{http://arxiv.org/abs/1312.0289}{arXiv:1312.0289}.
\bibitem[{Bo\.zek(2012)}]{Bozek:2011ua}
\bibinfo{author}{P.~Bo\.zek}, \bibinfo{journal}{Phys. Rev.}
  \bibinfo{volume}{C85} (\bibinfo{year}{2012}) \bibinfo{pages}{034901}.
  \DOIprefix\doi{10.1103/PhysRevC.85.034901}.
  \href{http://arxiv.org/abs/1110.6742}{arXiv:1110.6742}.
\bibitem[{Bo{\.z}ek and Broniowski(2014)}]{Bozek:2014cya}
\bibinfo{author}{P.~Bo{\.z}ek}, \bibinfo{author}{W.~Broniowski},
  \bibinfo{journal}{Phys.Lett.} \bibinfo{volume}{B739} (\bibinfo{year}{2014})
  \bibinfo{pages}{308--312}. \DOIprefix\doi{10.1016/j.physletb.2014.11.006}.
  \href{http://arxiv.org/abs/1409.2160}{arXiv:1409.2160}.
\bibitem[{Huang(2014)}]{phenixnapa}
\bibinfo{author}{S.~Huang} (\bibinfo{collaboration}{PHENIX Collaboration}),
  \bibinfo{journal}{talk given at the Workshop on Initial Stages of High Energy
  Nuclear Collisions, Napa, CA, December 3-7}  (\bibinfo{year}{2014}).
\bibitem[{Romatschke(2015)}]{Romatschke:2015gxa}
\bibinfo{author}{P.~Romatschke}  (\bibinfo{year}{2015}).
  \href{http://arxiv.org/abs/1502.04745}{arXiv:1502.04745}.
\bibitem[{Czy\.z and Maximon(1969)}]{Czyz:1969jg}
\bibinfo{author}{W.~Czy\.z}, \bibinfo{author}{L.~C. Maximon},
  \bibinfo{journal}{Annals Phys.} \bibinfo{volume}{52} (\bibinfo{year}{1969})
  \bibinfo{pages}{59--121}. \DOIprefix\doi{10.1016/0003-4916(69)90321-2}.
\bibitem[{Rybczy\'nski et~al.(2014)Rybczy\'nski, Stefanek, Broniowski, and
  Bo\.zek}]{Rybczynski:2013yba}
\bibinfo{author}{M.~Rybczy\'nski}, \bibinfo{author}{G.~Stefanek},
  \bibinfo{author}{W.~Broniowski}, \bibinfo{author}{P.~Bo\.zek},
  \bibinfo{journal}{Comput. Phys. Commun.} \bibinfo{volume}{185}
  (\bibinfo{year}{2014}) \bibinfo{pages}{1759}.
  \DOIprefix\doi{10.1016/j.cpc.2014.02.016}.
  \href{http://arxiv.org/abs/1310.5475}{arXiv:1310.5475}.
\bibitem[{Chojnacki et~al.(2012)Chojnacki, Kisiel, Florkowski, and
  Broniowski}]{Chojnacki:2011hb}
\bibinfo{author}{M.~Chojnacki}, \bibinfo{author}{A.~Kisiel},
  \bibinfo{author}{W.~Florkowski}, \bibinfo{author}{W.~Broniowski},
  \bibinfo{journal}{Comput. Phys. Commun.} \bibinfo{volume}{183}
  (\bibinfo{year}{2012}) \bibinfo{pages}{746}.
  \DOIprefix\doi{10.1016/j.cpc.2011.11.018}.
  \href{http://arxiv.org/abs/1102.0273}{arXiv:1102.0273}.
\bibitem[{Carlson and Schiavilla(1998)}]{Carlson:1997qn}
\bibinfo{author}{J.~Carlson}, \bibinfo{author}{R.~Schiavilla},
  \bibinfo{journal}{Rev. Mod. Phys.} \bibinfo{volume}{70}
  (\bibinfo{year}{1998}) \bibinfo{pages}{743}.
  \DOIprefix\doi{10.1103/RevModPhys.70.743}.
\bibitem[{Loizides et~al.(2014)Loizides, Nagle, and
  Steinberg}]{Loizides:2014vua}
\bibinfo{author}{C.~Loizides}, \bibinfo{author}{J.~Nagle},
  \bibinfo{author}{P.~Steinberg}  (\bibinfo{year}{2014}).
  \href{http://arxiv.org/abs/1408.2549}{arXiv:1408.2549}.
\bibitem[{Kharzeev and Nardi(2001)}]{Kharzeev:2000ph}
\bibinfo{author}{D.~Kharzeev}, \bibinfo{author}{M.~Nardi},
  \bibinfo{journal}{Phys. Lett.} \bibinfo{volume}{B507} (\bibinfo{year}{2001})
  \bibinfo{pages}{121--128}. \DOIprefix\doi{10.1016/S0370-2693(01)00457-9}.
  \href{http://arxiv.org/abs/nucl-th/0012025}{arXiv:nucl-th/0012025}.
\bibitem[{Back et~al.(2002)}]{Back:2001xy}
\bibinfo{author}{B.~B. Back}, et~al. (\bibinfo{collaboration}{PHOBOS}),
  \bibinfo{journal}{Phys. Rev.} \bibinfo{volume}{C65} (\bibinfo{year}{2002})
  \bibinfo{pages}{031901}.
  \href{http://arxiv.org/abs/nucl-ex/0105011}{arXiv:nucl-ex/0105011}.
\bibitem[{Bo\.zek and Wyskiel(2010)}]{Bozek:2010bi}
\bibinfo{author}{P.~Bo\.zek}, \bibinfo{author}{I.~Wyskiel},
  \bibinfo{journal}{Phys. Rev.} \bibinfo{volume}{C81} (\bibinfo{year}{2010})
  \bibinfo{pages}{054902}. \DOIprefix\doi{10.1103/PhysRevC.81.054902}.
  \href{http://arxiv.org/abs/1002.4999}{arXiv:1002.4999}.
\bibitem[{Bia\l{}as and Czy\.z(2005)}]{Bialas:2004su}
\bibinfo{author}{A.~Bia\l{}as}, \bibinfo{author}{W.~Czy\.z},
  \bibinfo{journal}{Acta Phys. Polon.} \bibinfo{volume}{B36}
  (\bibinfo{year}{2005}) \bibinfo{pages}{905}.
  \href{http://arxiv.org/abs/hep-ph/0410265}{arXiv:hep-ph/0410265}.
\bibitem[{Adare et~al.(2014)}]{Adare:2014keg}
\bibinfo{author}{A.~Adare}, et~al. (\bibinfo{collaboration}{PHENIX
  Collaboration})  (\bibinfo{year}{2014}).
  \href{http://arxiv.org/abs/1404.7461}{arXiv:1404.7461}.
\bibitem[{Adler et~al.(2002)}]{Adler:2002pu}
\bibinfo{author}{C.~Adler}, et~al. (\bibinfo{collaboration}{STAR
  Collaboration}), \bibinfo{journal}{Phys. Rev.} \bibinfo{volume}{C66}
  (\bibinfo{year}{2002}) \bibinfo{pages}{034904}.
  \DOIprefix\doi{10.1103/PhysRevC.66.034904}.
  \href{http://arxiv.org/abs/nucl-ex/0206001}{arXiv:nucl-ex/0206001}.
\bibitem[{Luzum and Ollitrault(2013)}]{Luzum:2012da}
\bibinfo{author}{M.~Luzum}, \bibinfo{author}{J.-Y. Ollitrault},
  \bibinfo{journal}{Phys. Rev.} \bibinfo{volume}{C87} (\bibinfo{year}{2013})
  \bibinfo{pages}{044907}. \DOIprefix\doi{10.1103/PhysRevC.87.044907}.
  \href{http://arxiv.org/abs/1209.2323}{arXiv:1209.2323}.
\bibitem[{Koop et~al.(2015)Koop, Adare, McGlinchey, and Nagle}]{Koop:2015wea}
\bibinfo{author}{J.~D.~O. Koop}, \bibinfo{author}{A.~Adare},
  \bibinfo{author}{D.~McGlinchey}, \bibinfo{author}{J.~Nagle}
  (\bibinfo{year}{2015}).
  \href{http://arxiv.org/abs/1501.06880}{arXiv:1501.06880}.
\bibitem[{Adare et~al.(2014)}]{Adare:2014vri}
\bibinfo{author}{A.~Adare}, et~al. (\bibinfo{collaboration}{PHENIX
  Collaboration})  (\bibinfo{year}{2014}).
  \href{http://arxiv.org/abs/1404.5291}{arXiv:1404.5291}.
\bibitem[{Kisiel et~al.(2006)Kisiel, Florkowski, Broniowski, and
  Pluta}]{Kisiel:2006is}
\bibinfo{author}{A.~Kisiel}, \bibinfo{author}{W.~Florkowski},
  \bibinfo{author}{W.~Broniowski}, \bibinfo{author}{J.~Pluta},
  \bibinfo{journal}{Phys. Rev.} \bibinfo{volume}{C73} (\bibinfo{year}{2006})
  \bibinfo{pages}{064902}. \DOIprefix\doi{10.1103/PhysRevC.73.064902}.
  \href{http://arxiv.org/abs/nucl-th/0602039}{arXiv:nucl-th/0602039}.
\bibitem[{Bo{\.z}ek(2014)}]{Bozek:2014fxa}
\bibinfo{author}{P.~Bo{\.z}ek}, \bibinfo{journal}{Phys.Rev.}
  \bibinfo{volume}{C90} (\bibinfo{year}{2014}) \bibinfo{pages}{064913}.
  \DOIprefix\doi{10.1103/PhysRevC.90.064913}.
  \href{http://arxiv.org/abs/1408.1264}{arXiv:1408.1264}.

\end{thebibliography}

\end{document}